%% file: COSINE_DAQ.tex
\documentclass[a4paper,11pt]{article}
\pdfoutput=1 % if your are submitting a pdflatex (i.e. if you have
\usepackage{jinstpub} % for details on the use of the package, please
\usepackage{lineno}
\usepackage{siunitx}

\title{\boldmath The COSINE-100 Data Acquisition System}

\input{authors}

\emailAdd{jayhyun.jo@yale.edu,jspark@post.kek.jp}

\abstract{
COSINE-100 is a dark matter direct detection experiment designed to test the annual modulation signal observed by the DAMA/LIBRA experiment. COSINE-100 consists of 8 NaI(Tl) crystals with a total mass of 106~kg, a 2200~L liquid scintillator veto, and 37 muon detector panels. We present details of the data acquisition system of COSINE-100, including waveform storage using flash analog-to-digital converters for crystal events and integrated charge storage using charge-sensitive analog-to-digital converters for liquid scintillator and plastic scintillator muon veto events. We also discuss several trigger conditions developed in order to distinguish signal events from photomultiplier noise events. The total trigger rate observed for the crystal/liquid scintillator (plastic scintillator) detector is 15 Hz (24 Hz). %check trigger rates
}


\collaboration{The COSINE-100 collaboration}
\begin{document}
\maketitle
%\linenumbers
\flushbottom
	
%\section{COSINE-100 Overview}
\section{Introduction}

The COSINE-100 experiment \cite{COSINE100:First} is a joint effort between KIMS-NaI~\cite{KIMS:2015, KIMS:2016, KIMS:2017} and DM-Ice~\cite{DMICE:2014, DMICE:2017} to unambiguously test the claim of dark matter detection by the DAMA/NaI~\cite{DAMA_NAI} and DAMA/LIBRA~\cite{DAMA_LIBRA:2010, DAMA_LIBRA:2013, DAMA_LIBRA2:2018} experiments with thallium-doped sodium iodide (NaI(Tl)) scintillating crystals. COSINE-100 is located at the Yangyang Underground Laboratory (Y2L), in the Yangyang pumped storage power plant, with 700~m minimum of rock overburden. The first phase of experiment with 106~kg of NaI(Tl) crystals began taking data in September 2016. 

\begin{figure}[!h]
	\begin{center}
		\includegraphics[scale=0.3]{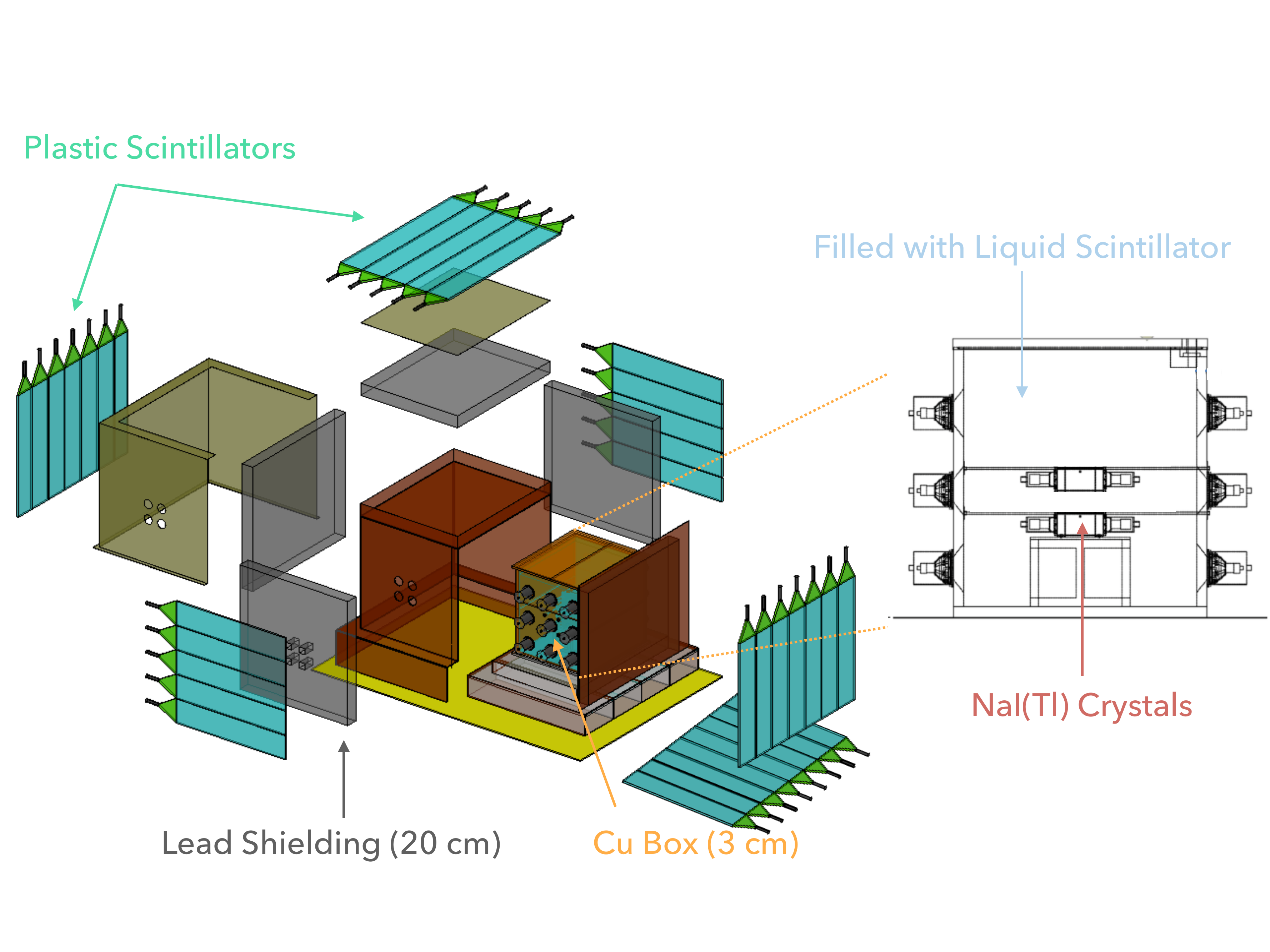}
	\end{center}
	\caption[]{A schematic of the COSINE-100 detector, which includes an NaI(Tl) crystal array with a total mass of 106~kg, a 2000~L liquid scintillator veto, copper and lead shielding, and 37 plastic scintillators for muon detection.} 
	\label{fig:cosine_detector}
\end{figure}

The COSINE-100 detector consists of eight NaI(Tl) crystals with a total mass of 106~kg, a 2200~L linear alkylbenzene (LAB)-based liquid scintillator (LS) veto for internal background suppression with coincidence event tagging~\cite{KIMS:LS} as well as cosmic-ray muons, copper and lead shielding, and 37 plastic scintillator (PS) panels that serve as an additional veto for cosmic-ray muons. Figure~\ref{fig:cosine_detector} shows a schematic representation of the COSINE-100 detector and shielding structure. A more detailed description of the detector can be found in ~\cite{COSINE100:First}.

A total of 16 3-inch photomultiplier tubes (PMTs) are attached to the crystals, with one PMT at each end of each crystal. The LS is contained in an acrylic-lined copper box, with 18 5-inch PMTs optically coupled to the acrylic box to detect photons from the LS veto. A total of 42 2-inch PMTs are attached to the plastic scintillators, two PMTs on each end of the five top panels and a single PMT for the rest. 

The COSINE-100 data acquisition system (DAQ) collects data from all detectors, including the NaI(Tl) crystals, liquid scintillator veto, and plastic scintillator veto. In this report, details of data acquisition system hardware, trigger algorithms for different data taking conditions, and the data flow are presented.

\begin{figure}[!h]
	\begin{center}
		\includegraphics[width=1.0\textwidth]{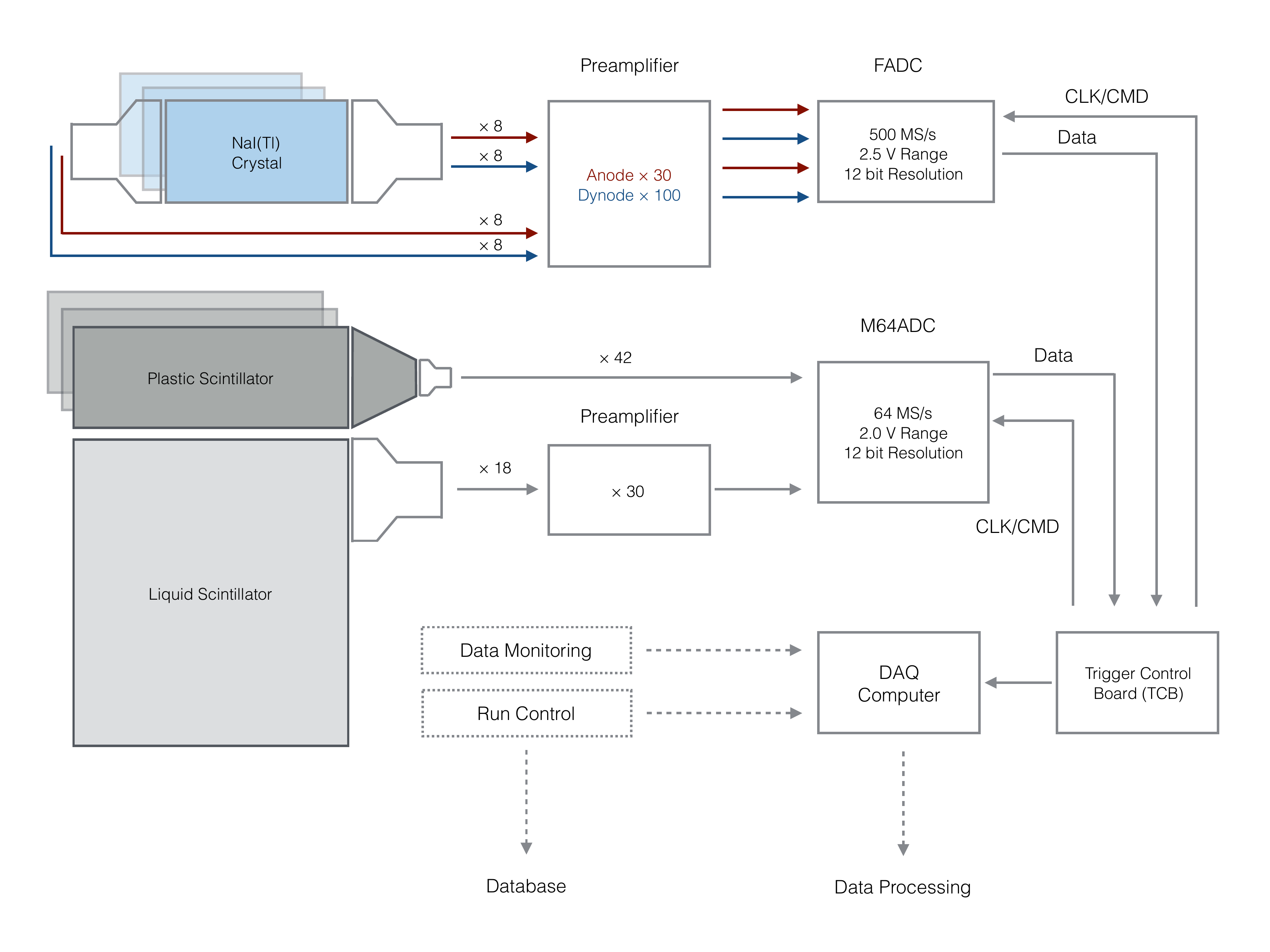}
	\end{center}
	\caption[]{The COSINE-100 data flow diagram. A total of 76 PMTs send signals to two types of digitizers that are managed by a trigger control board. The digitized data are saved in the DAQ computer. A total of 32 FADC channels and 60 M64ADC channels are used. InfluxDB and Grafana are used for storing and visualizing the data.} 
	\label{fig:data_flow}
\end{figure}

\section{Data Acquisition Modules and Trigger Conditions} \label{sec:daqmodules}

The data acquisition system of COSINE-100 consists of eight flash analog-to-digital converter (FADC\footnote{NKFADC500 module from NOTICE Korea: http://www.noticekorea.com}) modules, two charge-sensitive flash analog-to-digital converter (M64ADC\footnote{M64ADC module from NOTICE Korea}) modules, one trigger \& clock board (TCB), four preamplifiers, a high voltage (HV) supply system, and Linux processors. 

Liquid scintillator signals are amplified by a factor of 30 and fed into M64ADC modules, while the plastic scintillator signals are directly connected to M64ADCs (see Section~\ref{sec:m64adc}). Each of the 16 crystal PMTs provides two readouts (see Section~\ref{sec:pmt}), a high-gain signal from the anode and a low-gain signal from the 5th stage dynode, that are digitized by FADC modules (see Section~\ref{sec:fadc}). The anode signal is amplified by a factor of 30 and the dynode signal is amplified by a factor of 100. The 32 crystal signals, plus the 18 LS and 42 plastic scintillator PMTs amount to a total of 92 signal channels that are recorded by the DAQ system. The global trigger decisions are made by the TCB, which also synchronizes all of the module clocks (See Section~\ref{sec:tcb}); the raw data are stored on the DAQ computer in ROOT format. Figure~\ref{fig:data_flow} summarizes the overall data flow of the COSINE-100 experiment.

\subsection{Charge-Sensitive M64ADC for Liquid Scintillator and Muon Veto} \label{sec:m64adc}

To collect data from the LS and plastic scintillator vetoes, two charge-sensitive M64ADC modules with a 64~MS/s sampling rate are used. Each M64ADC module has 32 channels with an input dynamic range of 2~V$_{pp}$ input and 12-bit resolution. A total of 60 PMTs, 18 from the LS and 42 from the PS, are connected to these modules with BNC-type input connectors. The 18 PMT signals from the LS veto detector are amplified by a factor of 30, while the 42 PMT signals from the PS muon panels are not amplified but are directly connected to M64ADCs. The local trigger information from the ADC modules is passed to the TCB where a global trigger decision is made, and the final data is transferred to the DAQ computer via a USB3 port. 

\begin{figure}[]
	\begin{center}
		\includegraphics[width=0.7\textwidth]{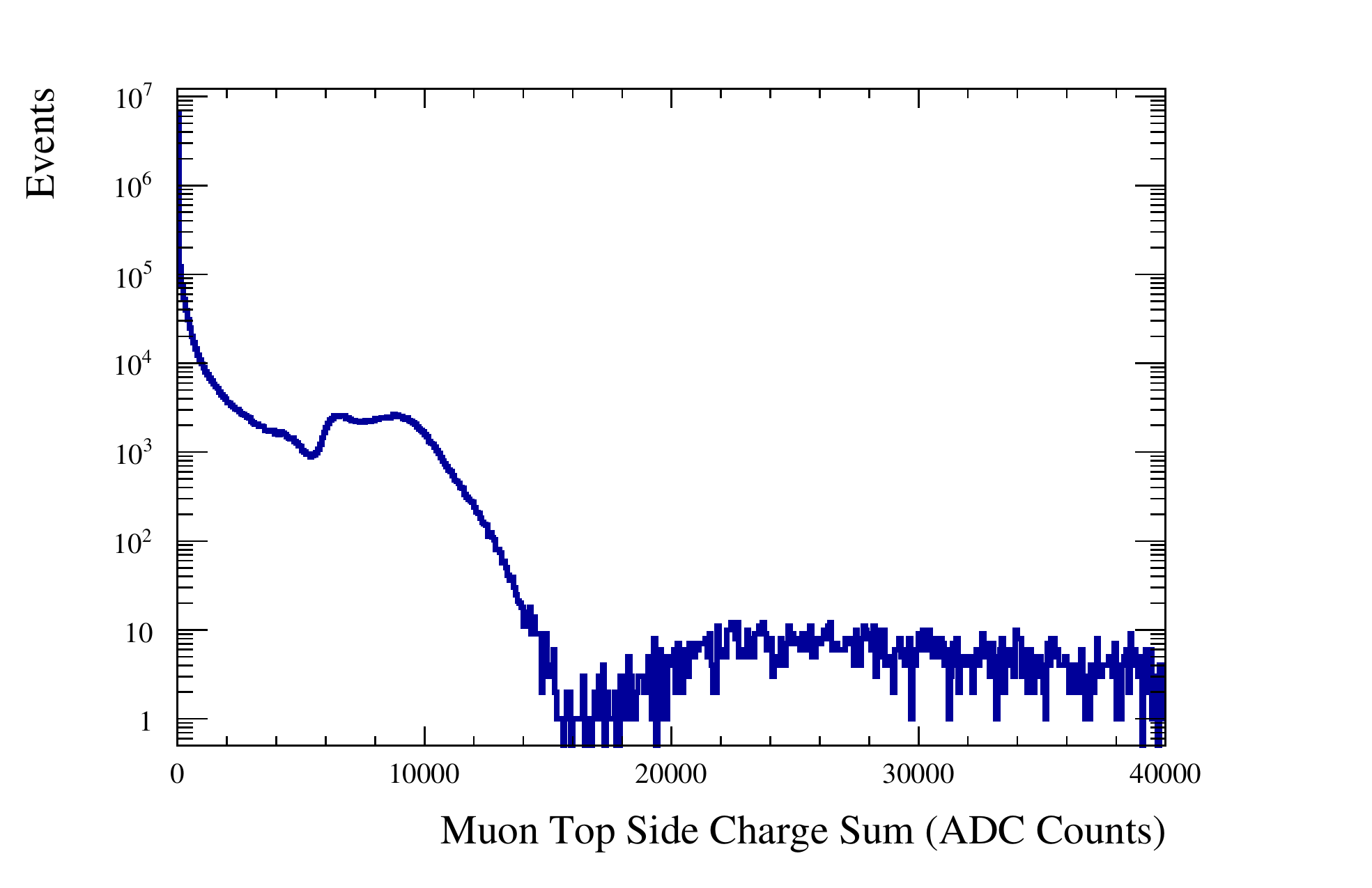}
	\end{center}
	\caption[]{The charge spectrum of one of the plastic scintillation panels for a live time of 120 hours. The charges from the two readout PMTs are summed and used to tag cosmic-ray muons. Events with integrated charges higher than 15,000~ADC counts are primarily due to muons, while lower energy events are mostly caused by Compton scattered $\gamma$-rays.} 
	\label{fig:muoncharge}
\end{figure}

The charge-sensitive M64ADCs do not save the raw waveform data; instead, they provide the signal time and integrated charge according to a preset field programmable gate array (FPGA) setting. The analog signal input to the M64ADC is digitized every 16~ns and, when the FPGA receives a trigger, it digitizes the input signal integrated over the following 192~ns, returning the charge sum in ADC counts.This 192~ns M64ADC module integration time is longer than the decay time of the plastic scintillator and the LAB-based organic scintillator. Figure~\ref{fig:muoncharge} displays an example spectrum of one of the top plastic scintillation panels.  %definintion of fast/slow component?

To select muon events and remove $\gamma$ background events caused by $\gamma$-rays that originate from outside the panels, we use an integrated charge trigger threshold of 4000~ADC counts (equivalent to approximately 125~pC); the integrated charge for most of the muon events is larger than 12,000~ADC counts (15,000~ADC counts for the top panels)~\cite{COSINE:Muon}. For the top panels that have two readout PMTs, either one of the two PMTs is required to have an integrated charge of more than 4000~ADC counts to be triggered. When a muon traverses the COSINE-100 detector, it deposits energy in at least two different muon detector panels and, depending on its trajectory, the LS veto as well. In the cases where a muon stops inside the LS, it will deposit energy in a single muon detector panel and the LS veto. Therefore, we categorize two classes of events: those with at least two hit muon panels or at least one hit muon panel and one LS-veto PMT. When the integrated charge of a single channel is larger than the preset threshold, a coincidence window with a 400~ns width is opened. If any other channel triggers within this window, the M64ADC sends a trigger signal to the TCB. See Section~\ref{sec:tcb} for more details on the TCB. 

\begin{figure}[]
	\begin{center}
		\includegraphics[scale=0.4]{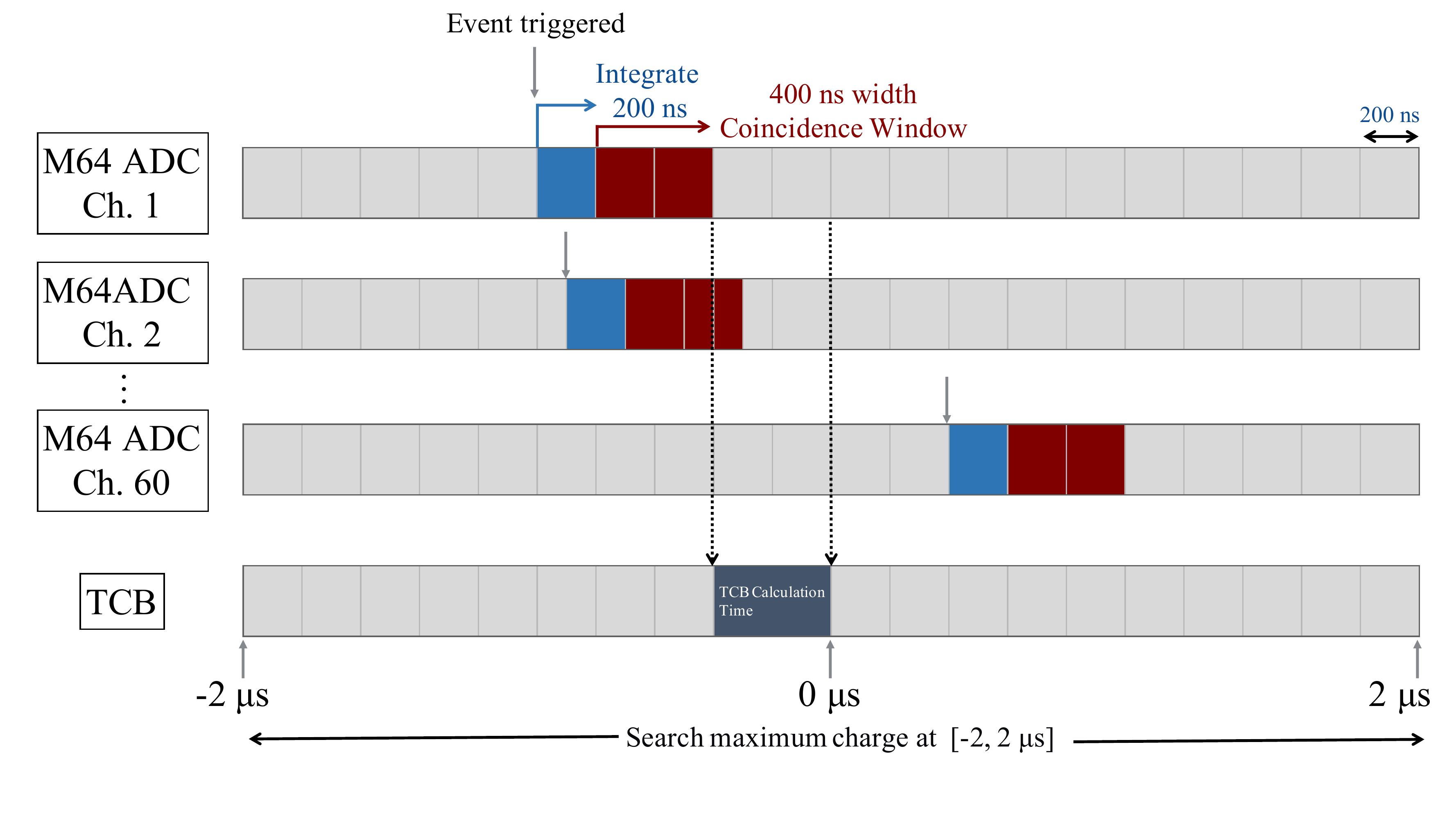}
	\end{center}
	\caption{Trigger sequence for the M64ADC system. We require coincidence of at least two PMT signals that occur within 400 ns from either the LS veto or muon detector panels. Blue boxes represent the charge integration time and red boxes indicates the 400~ns-wide coincidence window. TCB time indicates the time needed for the TCB to interpret information received from the M64ADCs and generate a global trigger.} 
	\label{fig:trigger_logic_m64}
\end{figure}

When the TCB sends a signal to the FADCs and M64ADCs to save an event, the M64ADCs open a \SI{4}{\micro\second} gate window, with the end of the TCB time in the middle of the window. Within this gate window, the M64ADC saves the maximum integrated charge values and its corresponding time for each channel. This time information is used for a passive data taking  algorithm that is described in Section~\ref{sec:tcb}. Figure~\ref{fig:trigger_logic_m64} illustrates the trigger timing sequence for the M64ADC system. %complicated...simpler/clearer explanation?

\subsection{Flash Analog-To-Digital Converter for Crystal Detectors} \label{sec:fadc}

A total of eight 500~MS/s FADC modules are used to digitize analog signals from the NaI(Tl) crystals. Each FADC has four input channels with SMA-type connectors, with each channel having a dynamic range of 2.5~V$_{pp}$ and a 12-bit resolution. Several functions are encoded in a FPGA on each FADC to define a local trigger for the module. 

All COSINE-100 crystals have a 3-inch PMT attached to each end. The anode signal, which is sensitive at the single-photoelectron level, is the primary channel for the dark matter search analysis. The dynode signal records high energy events, such as those from alpha particles, which are used to improve the understanding of the internal backgrounds in the crystal such as the U, Th, and Pb contamination. Because the PMTs that are used (Hamamatsu R12669SEL) are known to suffer from non-linearity when the signal energy is higher than 1~MeV\cite{KIMS:2015}, the 5th dynode is used for signals above this energy. The PMT base structure was modified to extract both anode and the 5th dynode signals and connect them to separate FADC channels. A detailed discussion of this non-linearity is provided in Section~\ref{sec:pmt}. 
For anode signals, a factor of 30 preamplifier is used to increase gain, with a linear response for energies up to 100~keV. A factor of 100 preamplifier is used for dynode signals and these have a linear energy response up to 3~MeV. %At each FADC, we connected each anode and dynode signal from both PMTs of each crystal. 

The waveforms from each of the two PMTs are recorded when there is a coincidence of high-gain anode signal larger than the preset threshold (6 mV, equivalent to 10~ADC counts or 0.2 photoelectrons) within a time window of 200~ns. The FADC module generates a trigger output signal if and only if both PMTs are triggered within the time window, as determined by the anode signal. The dynode signals are recorded when a global trigger is issued by the TCB. If any one of the eight crystals matches the coincidence condition, then a trigger signal is sent to the TCB (see Section~\ref{sec:tcb}) and the FADCs save the waveforms of all the channels. The recorded waveform is \SI{8}{\micro\second} long, starting approximately \SI{2.4}{\micro\second} before the trigger occurs. The pre-trigger information is stored in DRAM on the FADC board, which can store a waveform with up to \SI{64}{\micro\second} length. 

For FADCs, if a given PMT anode channel does not generate a trigger and exhibits only baseline fluctuations, the contents of waveforms for both anode and dynode channels are suppressed to zero. This zero suppression algorithm is critical to reduce the data size, as when the DAQ is triggered typically only one or two crystals have generated signals whereas all other channels contain only baseline data. With this zero suppression, the data size can be reduced by $\sim$80\%. 

\begin{figure}[]
	\begin{center}
		\includegraphics[width=0.8\textwidth]{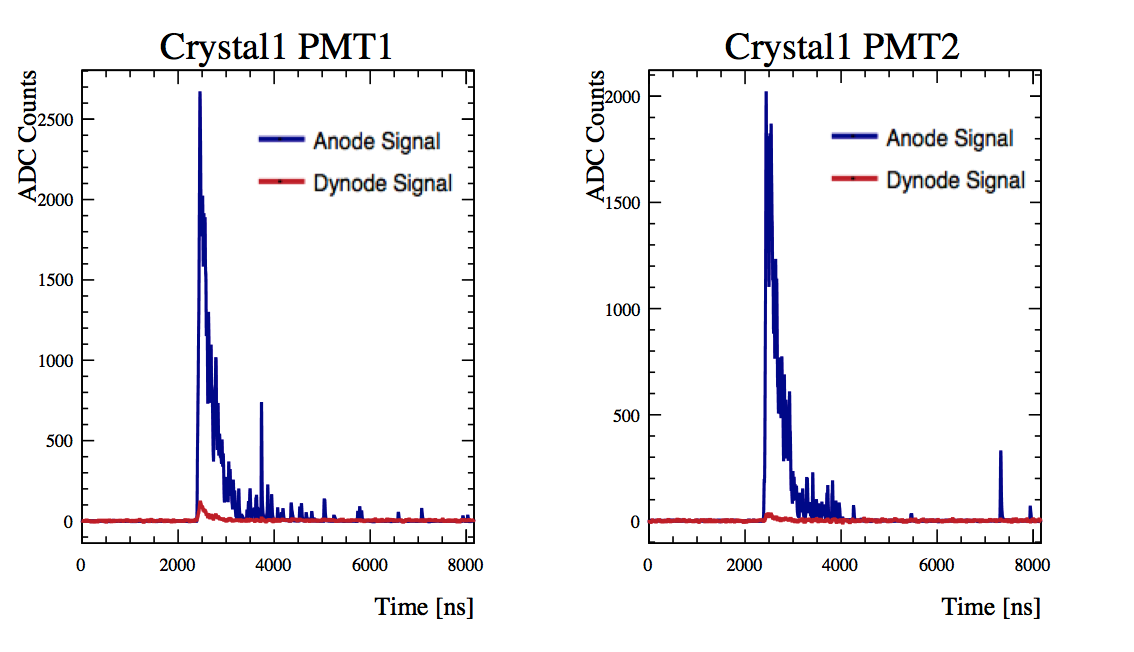}
	\end{center}
	\caption{Typical raw waveform of a triggered event from two PMTs in the same crystal, with both high-gain (anode) and low-gain (dynode) signals overlaid. The starting position of the waveform is set to \SI{2.4}{\micro\second} before a global trigger time.} 
	\label{fig:c1waveform}
\end{figure}

To minimize noise from the electronics, we have bare ground cables wrapped around the DAQ and HV SHV/BNC connectors, which results in a noise reduction for the readouts. The peak-to-peak noise level is $\sim$8~ADC counts on average ($\sim$1.4~ADC counts or $\sim$0.84~mV RMS) for the anode, and $\sim$20~ADC counts ($\sim$2.3~ADC counts or $\sim$1.38~mV RMS) for the dynode. Figure~\ref{fig:c1waveform} shows typical waveforms of a triggered signal event from two PMTs in a crystal, with both anode and dynode readouts. This specific event corresponds to an energy of approximately 60~keV.

%\begin{figure}[]
%	\begin{center }
%		\includegraphics[width=0.8\textwidth]{figs/C1_pedestal_new.png}
%	\end{center}
%	\caption{Pedestal waveform of a triggered event from two PMTs in the same crystal, with both high-gain (anode) and low-gain (dynode) signal overlaid.} 
%	\label{fig:c1pedestal}
%\end{figure}

\subsection{Trigger Condition, Trigger \& Clock Board, and Dead Time}\label{sec:tcb}

One of main roles of the LS veto is to tag internal background events originating from the crystals. If a crystal contains internal radioactive isotopes, such as $^{40}$K, $\gamma$ rays from decay products can deposit energy in a crystal and in the LS veto. It is particularly important to tag the $^{40}$K background in COSINE-100, as the $^{40}$K decay has a 3~keV X-ray line that is in the 2--6 keV energy range where the DAMA experiment observed an annual modulation~\cite{DAMA_NAI, DAMA_LIBRA:2010,DAMA_LIBRA:2013}. This background can be suppressed by tagging coincident events of the 3~keV X-ray in the crystal and an accompanying 1460~keV $\gamma$-ray in the LS or the other crystals.

\begin{figure}[]
	\begin{center}
		\includegraphics[scale=0.4]{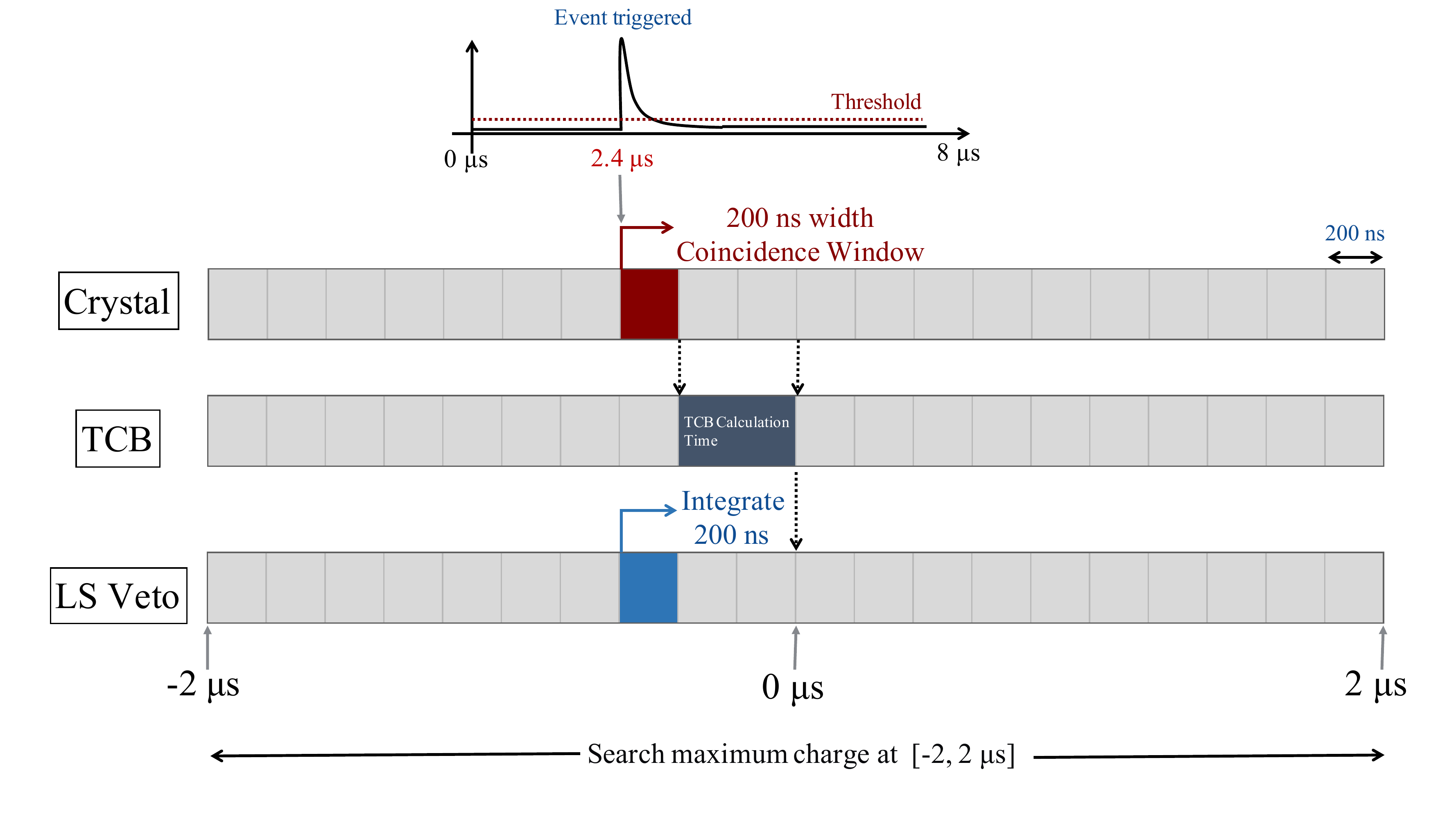}
	\end{center}
	\caption[]{Trigger algorithm for the case of passive data taking. Blue boxes represent the charge integration time and red boxes indicate the 200~ns wide coincidence windows. TCB time indicates the time needed for the TCB to interpret information received from the FADCs and M64ADCs and then generate a global trigger. We require coincident signals in both PMTs on a single crystal to occur within 200~ns of each other. If the trigger condition is satisfied, \SI{8}{\micro\second} of crystal waveform and M64ADC data for a single event is collected at the same time.} 
	\label{fig:passive_data}
\end{figure}

\begin{figure}[]
	\begin{center}
		\includegraphics[width=0.7\textwidth]{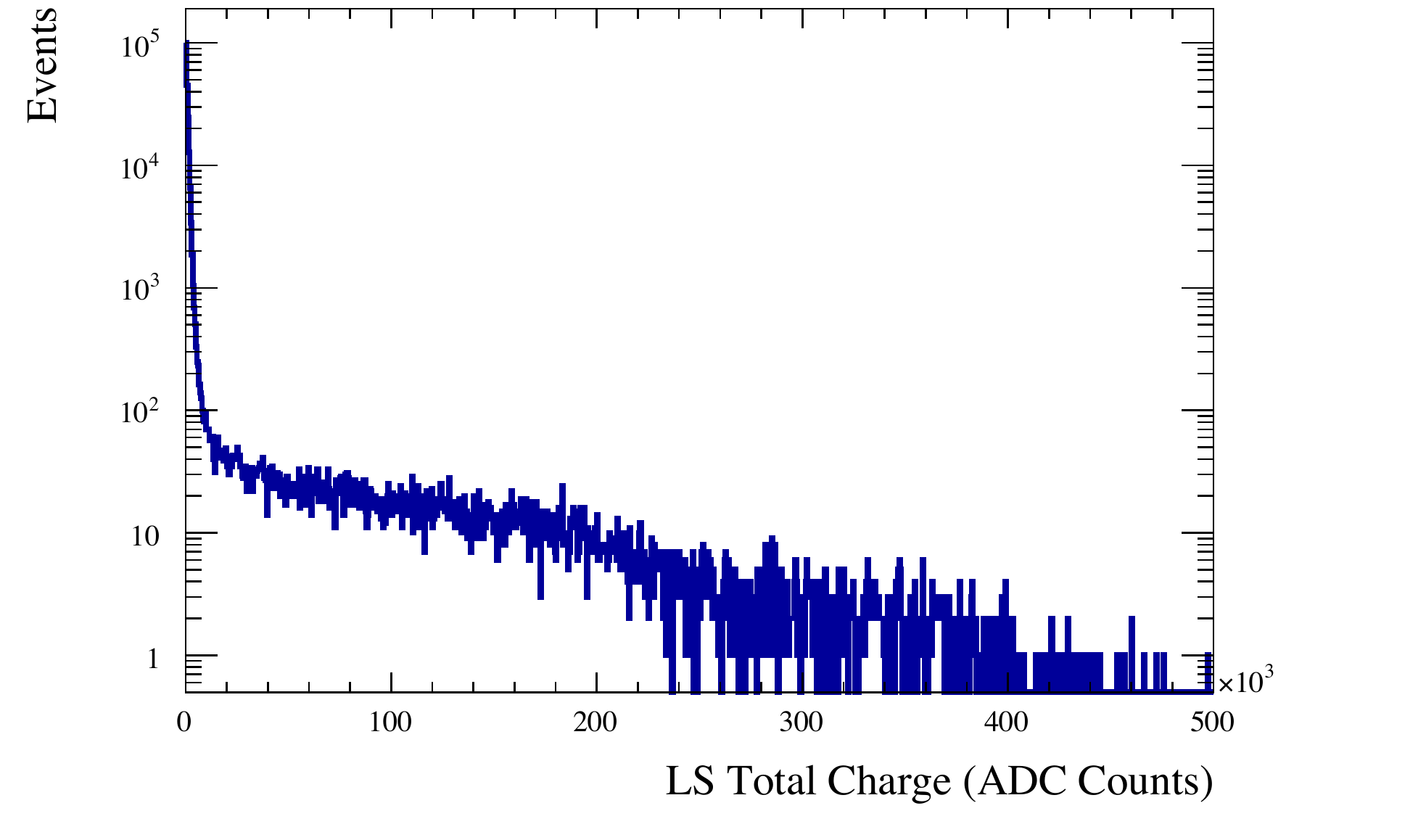}
	\end{center}
	\caption[]{Liquid scintillator charge spectrum with two hours of crystal trigger. The charge from all 18 LS PMTs are summed.} 
	\label{fig:lscharge}
\end{figure}

The trigger signals generated from the FADCs or charge-sensitive M64ADCs are sent to the TCB, where a global trigger is created. There are two types of global triggers: The first is when there is a trigger from the FADC, which results in all the channels including the FADC (crystal) and the M64ADC (both the plastic and liquid scintillator veto) being recorded. Here the M64ADC saves the data when the FADC generates a trigger, and is thus termed "passive data taking". Figure~\ref{fig:passive_data} describes the trigger algorithm for the passive data taking and Fig.~\ref{fig:lscharge} shows the total integrated charge of the LS veto with two hours of such passive data taking.

For the second type of trigger, the M64ADC modules corresponding to the plastic scintillator veto are actively triggered when the trigger condition (multiplicity and threshold) is met. The FADC channels are not necessarily recorded in this case. This type of the global trigger is explained in Sec.~\ref{sec:m64adc} and in Fig.~\ref{fig:trigger_logic_m64}.

These two global trigger paths between the FADC and M64ADC modules corresponding to the plastic scintillator veto are mutually independent and they both use multiplicity for deciding a global trigger; for M64ADC modules corresponding to the liquid scintillator veto, there is no independent trigger path as they are only passively triggered by crystal or  plastic scintillator veto.

The global trigger is sent back to all FADCs and/or charge-sensitive M64ADCs at the same time depending on the type of trigger. The TCB has 40 RJ-45 type slots and the FADCs and M64ADCs are distinguished by slot number; and the communication between the TCB and FADCs or M64ADCs is done via Ethernet cables. The TCB also sends the TCB clock count to the modules which is used to synchronize the event time across modules. 

When muons pass through a NaI(Tl) crystal, they cause a very high trigger rate (greater than 1~kHz) which exceeds the capacity of the DAQ system. Both the KIMS-NaI~\cite{KIMS:2015} and DM-Ice experiments~\cite{DMICE:Muon} have measured elevated trigger rates of muon-induced events within NaI(Tl) targets. To prevent this, we apply a 1~ms hardware dead time to the FADC channel that generates a trigger. Because the the DAQ rate of the FADCs is about 15~Hz, the total dead time is only about 1.5\%.

\section{Other Electronics}\label{sec:others}

\subsection{Preamplifier}\label{sec:preamp}

The preamplifier is designed to amplify the single photoelectron signals of a PMT without any loss of pulse information. As described in Section.~\ref{sec:m64adc} and \ref{sec:fadc}, signals from crystal PMTs are amplified by a factor of 30 (anode) and 100 (dynode); additionally, the LS PMT signal is amplified by a factor of 30. Customized preamplifiers with BNC connectors are used for these amplifications. The bandwidth of a PMT signal is about 100~MHz which requires a sampling rate greater than 200~MHz, which led us to use a 200~MHz bandwidth op-amp to amplify the signal. 

%The preamplifiers' BNC connector points are wrapped with bare ground cables in order to reduce noises by factor of 3 (5) for anode (dynode) signal from crystal and LS PMTs.

\subsection{High Voltage Supply Module} \label{sec:hv}

In order to supply high voltage to PMTs, CAEN high voltage supply modules are used. We use 8 CAEN A1535N HV supplies with two 24-channel and six 12-channel modules, which are all installed in a CAEN-4527 VME crate. All modules can supply voltages up to 3.5~kV with negative polarity, sufficient for all the PMTs used in the COSINE-100 detectors, as the highest magnitude of the voltage applied does not exceed 2500~V after the PMT gain correction. A high voltage scan was performed for all the channels prior to the physics run, followed by gain matching and light yield measurements\cite{COSINE100:First}. 

%All the HV modules' SHV connector points are wrapped with bare ground cables to reduce noise, just like the pre-amplifier.

\subsection{Photomultiplier Tubes}\label{sec:pmt}

The COSINE-100 experiment uses the 3-inch diameter R12669SEL photomultiplier tubes made by Hamamatsu Photonics, a modified version of the R6233 Super Bialkali (SBA) PMT used in DAMA~\cite{DAMA_LIBRA2:2018}, to detect the scintillation light from NaI(Tl) crystals. This light sensor is made from low-radioactivity material and exhibits a high quantum efficiency of 35\%, leading to enhanced sensitivity for low energy events below 10~keV. 

As descried in Section~\ref{sec:fadc}, signals from the anode and dynode of the PMTs attached to the NaI(Tl) crystals are taken simultaneously. To understand the internal backgrounds by looking at high energy alpha events, linearity over a wide energy range is important, but R12669SEL PMTs suffer from non-linear behavior at energies above a few MeV. 

%Specifically, there are two cases where a signal may have to be taken from a dynode: 1) when the measurement to be made requires a signal synchronous with the anode signal and 2) when it is desirable to limit gain by using fewer stages. %Procedures applicable to these cases are different. 
%The requirement for a synchronous signal from a dynode usually arises in connection with the detection of very short pulses. The signal may be required either for synchronizing an instrument or for supplying additional charge or amplitude information. The problem is to obtain the required signal without disturbing the anode signal. %Fig.~\ref{fig:Xdynode} shows that even though the signal is usualyy taken from the last dynode, but it's not needed to be if an amplitude is not a governing consideration, as it is still possible to obtain an amplitude comparable with the anode, with different stages of dynode.

%Fig.~\ref{fig:Xdynode} shows a solution. To obtain an amplitude comparable with that at the anode, the signal is usually taken from the last dynode but need not to be if amplitude is not a governing consideration.

%\begin{figure*}[!htb]
%	\begin{tabular}{cc}
%		\includegraphics[width=1.0\textwidth]{figs/pmt/dynode_pulse_height.eps}&
%	\end{tabular}
%	\caption{Amplification comparison of each dynode with anode pulses.}
%	\label{fig:Xdynode}
%\end{figure*}

%We have investigated the R12669SEL tube to understand the linearity at a wider energy range and observed that its high energy non-linearity can be resolved by reading out signals from one of the lower dynodes and the anode, synchronously.

\begin{figure*}[]
	\begin{center}
		\includegraphics[width=0.7\textwidth]{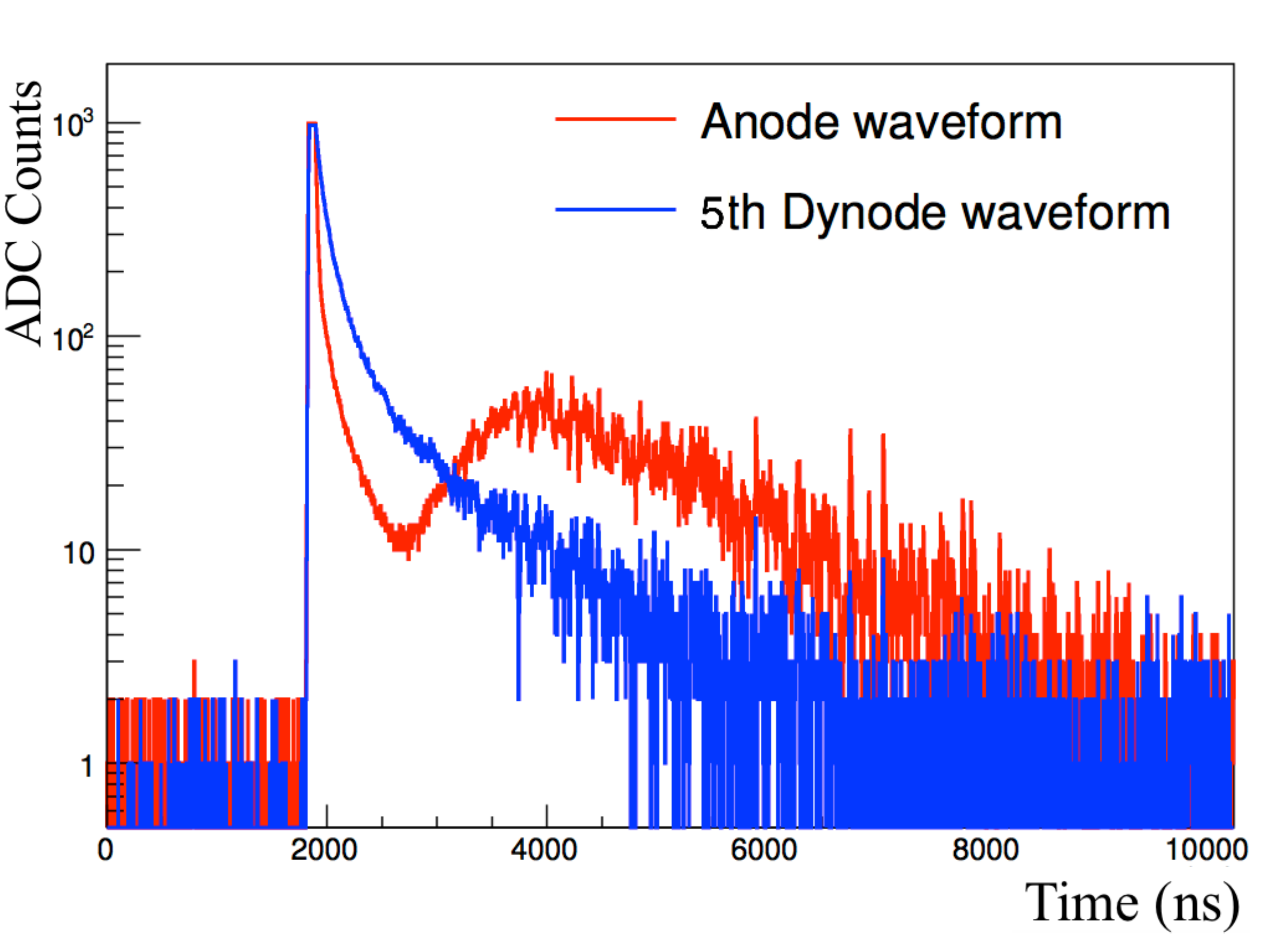}
	\end{center}
	\caption{Raw waveform of an event read out by both the anode and dynode-stage 5. The anode waveform shows a distortion due to the non-linearity, whereas the dynode-stage 5 readout of the same event shows a normal waveform.}
	\label{fig:anodeDynodeWaveform}
\end{figure*}

\begin{figure*}[]
	\begin{center}
		\includegraphics[width=1.0\textwidth]{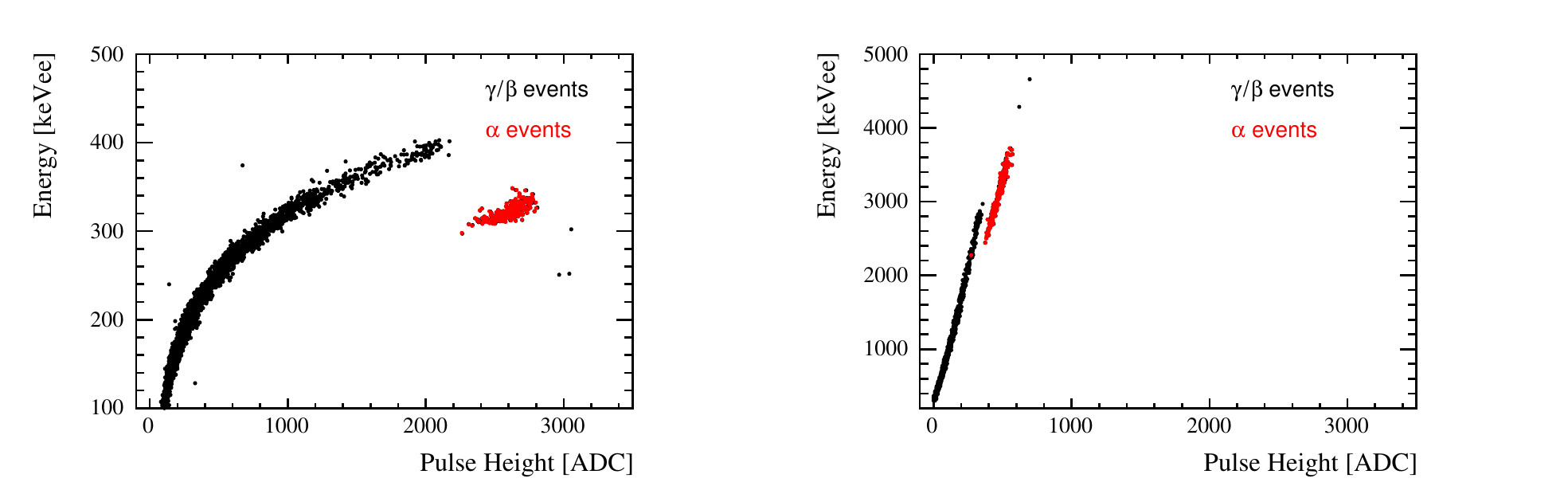}
	\end{center}
	\caption{Energy distribution as a function of pulse height for anode channel (left) and 5th dynode channel (right). The anode signal shows non-linearity for high energy, whereas the dynode signal has a much improved linearity. Red data points are high energy alpha events.}
	\label{fig:energyComparison}
\end{figure*}

The R12269SEL PMT shows two characteristic charge non-linearity behaviors: one is mainly due to saturation of a pulse while the other is due to electron loss at the last stages of dynodes. It is known that space charges created near the final dynode stages may cause some electrons not to reach the next stage. To resolve this issue, a test setup with a modified PMT base circuitry was developed with different stages of dynode channel readout. When the waveforms of the anode and the dynode-stage 5 are compared in Fig.~\ref{fig:anodeDynodeWaveform}, one can see that the waveform shape is distorted at the anode readout, but is preserved at the dynode-stage 5 readout. Figure~\ref{fig:energyComparison} also clearly shows a deformation in the correlation between charge sum and a pulse height in the anode channel, whereas in the dynode-stage 5 the linearity is maintained up to 1 MeV. 

\begin{figure*}[]
	\begin{center}
		\includegraphics[width=0.7\textwidth]{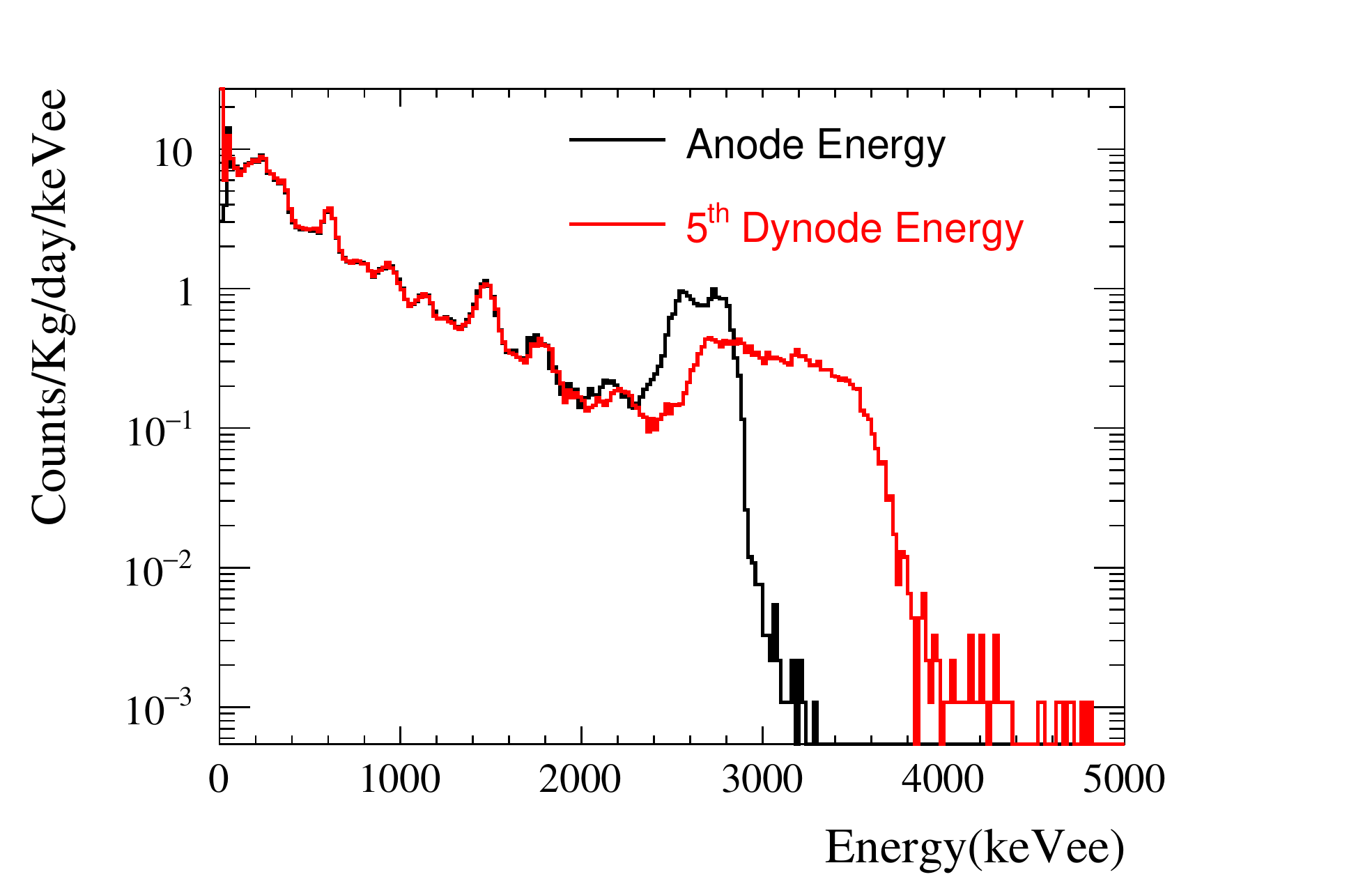}
	\end{center}
	\caption{Energy spectrum of the anode and the 5th dynode signal. The anode signal shows saturation in high energy where the dynode-stage 5 signal extends up to 4~MeV. The dynode signal also shows that it can maintain an energy resolution as good as the one from the anode signal up to few MeV.}
	\label{fig:anodDynodeSpectrum}
\end{figure*}

To check if the modified dynode channel can reproduce the anode signal with reasonable energy resolution, the energy spectrum of NaI(Tl) crystals has been studied with the test setup. Figure~\ref{fig:anodDynodeSpectrum} shows the background distribution of the dynode signal from stage 5 overlaid with the anode spectrum, illustrating that dynode signals can reproduce anode signals. Figure~\ref{fig:anodDynodeSpectrum} also indicates that the 5th dynode signal, while having a lower gain than the anode channel, still maintains an energy resolution as good as the one from the anode signal. 

\section{Data Quality Monitoring}\label{sec:monitoring}

\begin{figure}[]
	\begin{center}
		\includegraphics[width=0.7\textwidth]{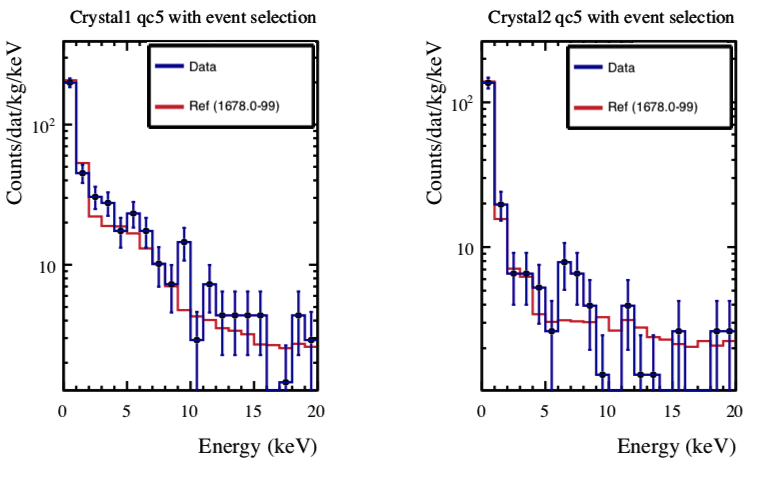}
	\end{center}
	\caption[]{Example plots from the online data monitoring system. This set of plots are displaying energy spectra of two crystals with a pre-selection of triggered events. Blue points are the data from the 2 hours long subrun file, and red lines represent reference plots from a "good" subrun data.}
	\label{fig:monitoring}
\end{figure}

To monitor the data quality, the COSINE-100 experiment has an online data monitoring system with a dedicated web server. The raw data are converted to a ROOT format and saved in the DAQ computer every two hours and the data are processed to extract monitoring variables. The monitoring system displays 26 variables corresponding to the FADC modules (one for total trigger rate, eight for crystal variables, and 17 for PMTs variables) and six corresponding to the M64ADC modules (one for total trigger rate, two for plastic scintillator PMTs, and three for liquid scintillator PMTs). All the variable plots are overlaid with reference plots, with known good data quality, to highlight possible deviations of the monitoring variables. The reference plot is frequently updated to account for time variation of the variables. Figure~\ref{fig:monitoring} shows example plots of one of the monitoring variables, the low energy spectrum with basic noise event removal. The blue line indicates data from the two-hour-long subrun file, whereas red lines represent the reference plots from good subrun data. Any deviation from the reference plot will be examined afterwards.

\begin{figure}[]
	\begin{center}
		\includegraphics[width=0.7\textwidth]{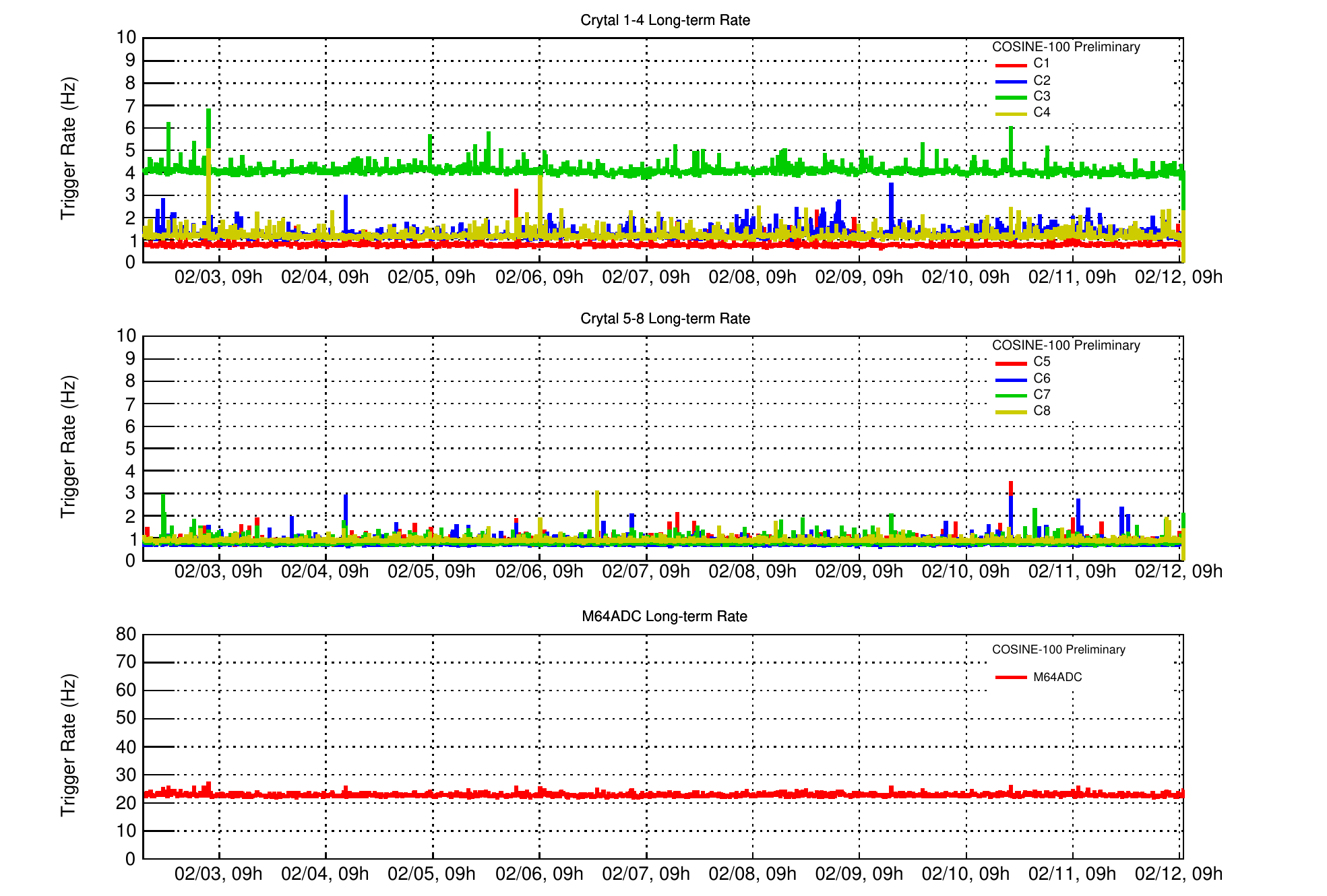}
	\end{center}
	\caption[]{FADC (Crystal) and M64ADC (veto) trigger rates as functions of time, between Feb.~2, 2018 and Feb.~12, 2018. All the crystals and M64ADC PMTs show stable rates during this period, without any abnormalities. A higher rate in C3 (Crystal-3) than in other crystals is observed to be due to higher PMT noise. This plot is generated by the COSINE-100 monitoring system (see Section~\ref{sec:monitoring}).}
	\label{fig:rates}
\end{figure}

In order to monitor longer-term data accumulation, a total of seven weekly monitoring plots are also generated automatically: live-time, trigger rates of all crystals, M64ADC trigger rate, LS total charge, LS charge asymmetry, and muon event rate. For muon event rate monitoring, a higher level muon event selection, that includes both the charge and timing information, is applied to identify muon events among PS triggered events. These plots help to identify any long term trends in the data, such as a gradual increase in event rate. Figure~\ref{fig:rates} shows event rates for all 8 crystals and M64ADC PMTs, from February 2, 2018 to February 12, 2018. 

%Fig.~\ref{fig:weeklymonitoring} is the example plot for the last 3 variables, from October 16th to 24th, 2017. These plots help to identify any abnormal behavior in data in a longer period time that's harder to notice in 2-hours subrun monitoring, such as a gradual and slow increase in event rate. 

%\begin{figure}[]
%	\begin{center}
%		\includegraphics[scale=0.4]{figs/monitoring_weekly.png}
%	\end{center}
%	\caption[]{Example plots from the weekly data monitoring. Three plots from M64ADC data are shown here, the LS total charge, the LS charge asymmetry, and the muon rate.}
%	\label{fig:weeklymonitoring}
%\end{figure}

\section{Summary}

The COSINE-100 detector was successfully constructed and commissioned by September 2016, and began its physics run in early October 2016. The data acquisition system for COSINE-100 is comprised of different ADC modules for the main crystal detectors and liquid or plastic scintillator vetoes. Trigger conditions and thresholds were carefully set to ensure stable data taking and effective coincident event selection, which is crucial for tagging the internal backgrounds of the crystals. 

The total trigger rate observed for the crystal (plastic scintillator) detectors is 15~Hz (24~Hz). The current live time of the detector is $>$91\%, where the $\sim$9\% down time originates primarily from calibrations and maintenance downtime.

\acknowledgments

We thank the Korea Hydro and Nuclear Power (KHNP) company for providing the underground laboratory space at Yangyang. This research was funded by the Institute for Basic Science (IBS) under project code IBS-R016-A1, Republic of Korea; an Alfred P. Slone Foundation Fellowship; NSF Granf Nos. PHY-1151795, PHY-1457995, DGE-1122492, and DGE-1256259; the Wisconsin Alumni Research Foundation; UIUC Campus Research Board; Yale University; and DOE/NNSA Grand No. DE-FC52-08NA28752, United States; STFC Grant ST/N000277/1, United Kingdom; and CNPq and grant \#2017/02952-0, S\~ao Paulo Research Foundation (FAPESP), Brazil.

\end{document}

%% file: authors.tex
\author[a]{G.~Adhikari,}
\affiliation[a]{Department of Physics, Sejong University, Seoul 05006, Republic of Korea}
\author[a]{P.~Adhikari,}

\author[b]{E.~Barbosa de Souza,}
\affiliation[b]{Wright Laboratory, Department of Physics, Yale University, New Haven, Connecticut 06520, USA}

\author[c]{N.~Carlin,}
\affiliation[c]{Physics Institute, University of S\~{a}o Paulo, 05508-090, S\~{a}o Paulo, Brazil}
\author[d]{S.~Choi,}
\affiliation[d]{Department of Physics and Astronomy, Seoul National University, Seoul 08826, Republic of Korea} 
\author[e]{W.~Choi,}
\affiliation[e]{Korea Institute of Science and Technology Information, Daejeon, 34141, Republic of Korea }

\author[f]{M.~Djamal,}
\affiliation[f]{Department of Physics, Bandung Institute of Technology, Bandung 40132, Indonesia}

\author[g]{A. C. ~Ezeribe,}
\affiliation[g]{Department of Physics and Astronomy, University of Sheffield, Sheffield S10 2TN, United Kingdom}

\author[h]{C.~Ha,}
\affiliation[h]{Center for Underground Physics, Institute for Basic Science (IBS), Daejeon 34126, Republic of Korea}
\author[i]{I.S.~Hahn,}
\affiliation[i]{Department of Science Education, Ewha Womans University, Seoul 03760, Republic of Korea} 
\author[b]{A. J. F.~Hubbard,}

\author[h]{E.J.~Jeon,}
\author[b]{J. H.~Jo}
\author[d]{H.W.~Joo,}

\author[h]{W.G.~Kang,}
\author[j]{W.S.~Kang,}
\affiliation[j]{Department of Physics, Sungkyunkwan University, Seoul 16419, Republic of Korea}
\author[k]{M.~Kauer,}
\affiliation[k]{Department of Physics and Wisconsin IceCube Particle Astrophysics Center, University of Wisconsin-Madison, Madison, 53706, USA}
\author[h]{H.~Kim,}
\author[l]{H.J.~Kim,}
\affiliation[l]{Department of Physics, Kyungpook National University, Daegu 41566, Republic of Korea}
\author[d]{K.W.~Kim,}
\author[j]{M.C.~Kim,}
\author[h]{N.Y.~Kim,}
\author[d]{S.K.~Kim,}
\author[a,h]{Y.D.~Kim,}
\author[h,m]{Y.H.~Kim,}
\affiliation[m]{Korea Research Institute of Standards and Science, Daejeon 34113, Republic of Korea}
\author[g]{V. A.~Kudryavtsev,}

\author[h]{H.S.~Lee,}
\author[h]{J.~Lee,}
\author[h]{M.H.~Lee,}

\author[b]{K. E.~Lim,}
\author[h]{D.S.~Leonard,}

\author[b]{R. H.~Maruyama,}
\author[g]{F.~Mouton,}

\author[h]{S.~Na,}

\author[h]{S.L.~Olsen,}

\author[h]{H.K.~Park,}
\author[m]{H.S.~Park,}
\author[h,o]{J.S.~Park}
\author[h]{K.S.~Park,}

\author[b]{W.~Pettus,}
\author[f]{H.~Prihtiadi,}

\author[j]{C.~Rott,}

\author[g]{A.~Scarff,}
\author[g]{N. J. C.~Spooner,}

\author[b]{W. G.~Thompson,}

\author[n]{L.~Yang}
\affiliation[n]{Department of Physics, University of Illinois at Urbana-Champaign, Urbana, Illinois 61801, USA}

\affiliation[o]{\textit{Present address}: High Energy Accelerator Research Organization (KEK), Ibaraki 319-1106, Japan}

\date{\today}